# Four Decades of Digital Waveguides

**PABLO TABLAS DE PAULA,**[1] *AES Member,* **JULIUS O. SMITH III,**[2] *AES Fellow,*
**VESA VÄLIMÄKI,**[3] *AES Fellow,* **AND JOSHUA D. REISS,**[1] *AES Fellow*

(p.tablasdepaula@qmul.ac.uk)   (jos@ccrma.stanford.edu)   (vesa.valimaki@aalto.fi)   (joshua.reiss@qmul.ac.uk)

[1]*Centre For Digital Music (C4DM), Queen Mary University of London, London, United Kingdom*
[2]*Center for Computer Research in Music and Acoustics (CCRMA), Stanford University, Stanford, CA, USA*
[3]*Acoustics Lab, Dept. Information and Communications Engineering, Aalto University, Espoo, Finland*

*Digital waveguide* physical modeling offers efficient simulation of acoustic wave propagation as compared to general finite-difference schemes commonly used in computational physics. This efficiency has enabled the real-time implementation of physically modeled musical instruments and sound effects, as well as real-time vocal models and artificial reverberation. This paper provides an overview of the historical evolution and applications of digital waveguide modeling and highlights recent advances in the field. Parametric optimization using classical, evolutionary and neural approaches are also discussed and compared. Digital waveguides provide physically accurate simulations with reduced computational cost, and can now be optimized with modern machine learning and differentiable digital signal processing techniques.

## 0 INTRODUCTION

Since their introduction forty years ago, *digital waveguide* (DWG) models have become a *cornerstone* of physically based sound synthesis and processing, spanning musical-instrument modeling, sound effects, artificial reverberation, and singing-voice synthesis [1–5]. Here we review four decades of DWG research, drawing together their historical evolution, theoretical foundations, and modern applications.[1]

In musical sound synthesis, physics-based equations are rarely used, even when the goal is to create realistic and expressive models of real-world instruments and effects. Instead, the prevailing approach often involves making numerous recordings that span the desired range of sounds—a process that is both labor-intensive and memory-hungry. Historically, this preference arose because solving the differential equations of physics by means of finite-difference schemes was computationally prohibitive. Although advances in methods have made physical modeling more feasible, further acceleration remains a valuable objective.

The digital-waveguide approach to acoustic modeling tackles this challenge by optimizing the physical computations for nearly lossless one-dimensional wave-propagation media, such as vibrating strings, acoustic tubes, and woodwind bores. In the frequency domain, all such systems exhibit quasi-*harmonic* resonant modes. The method has also been extended beyond one dimension through *waveguide meshes*. A rectangular lattice of 1D waveguides can approximate a 2D membrane, enabling efficient simulation of membranes and plates, with $N$-dimensional extensions following similarly.

The rest of this article is organized as follows. §1 charts key developments from the 1985 reverberator prototype to contemporary DWG-based products, comparing to related methods. §2 lays out the foundations and building blocks of DWGs. §3 surveys major extensions, and §4 surveys methods to optimize the DWG parameters, from classical physics-based, filter-design and system-identification techniques to genetic, modern machine learning and differentiable digital signal processing techniques.

---

[1]Interactive audio examples and supplementary materials are available at `https://joshreiss.github.io/digital-waveguides-review/`, The full source code is openly accessible, and the demos synthesize audio in real-time within any modern web browser using the Web Audio API.

## 1 HISTORICAL BACKGROUND

Digital waveguides were introduced in the 1985 paper 'A new approach to digital reverberation using closed waveguide networks' [6]. The backstory is that Gary Kendall was working on large artificial reverberation systems, which he discussed with Julius Smith on the way to a conference. Kendall's problem was that when he added a new feedback connection to enrich the response, the network would usually go unstable.

Since Schroeder and Logan in 1961 [7], there have been well-developed methods for using allpass filters to walk the boundary between stable and unstable systems for artificial reverberation. But there was not yet a general approach for adding arbitrary signal feedback paths without significantly affecting stability or overall reverberation time. An allpass construction approach was needed that supported any signal-path delay and interconnection topology. Smith pondered this while studying Belevitch [8] regarding allpass networks, and the idea of closed networks of lossless DWGs struck him as a promising solution.

In hindsight, DWGs can be regarded as a straightforward application of the following:

1. D'Alembert's *traveling-wave solution* of the wave equation [9];
2. *Scattering theory* implicit in d'Alembert's derivation and later developed for transmission lines;
3. *Sampling* of bandlimited signals and systems as taught by Whittaker, Nyquist, and Shannon [10].

However, it was many years after d'Alembert's solution that Lagrange calculated the first demonstration of reflection and transmission at the boundary of two ideal strings having different mass densities (1759), and scattering theory developed about a century later in optics (e.g., Fresnel 1823, Lord Rayleigh 1871). The concept of wave impedance awaited Heaviside's "telegrapher's equations" (coining "impedance" in 1886).

It can also be argued that standard physical-modeling formulations teach *away* from DWGs, because we normally begin with a *differential equation* giving constraints at every point in time and space, and we discretize that to make a generative model using *finite differences*. Since all practical models require *damping*, pure delay lines do not emerge from practical finite-difference schemes. One must come up with the *psychoacoustically valid approximation* of *summarizing* losses and dispersion at waveguide boundaries. Schroeder introduced delay-line simulation of sampled 1D wave propagation [11], extended by Moorer to include wall-reflection damping filters [12]. In the case of 1D linear, time-invariant waveguides, point-to-point transfer functions remain invariant as damping and/or dispersion filtering are *commuted* with delay elements to either endpoint, yielding an exact simplification; in higher dimensions, and for nonlinear or time-varying waveguides, it usually remains a high-quality approximation, despite the lack of commutative invariance.

In the early history of digital signal processing, there was a well developed technology of ladder and lattice digital filters that could be derived from sampled traveling-wave theory [13]. Along those lines, a digital singing voice having a ladder-filter vocal tract was demonstrated at Bell Labs by Kelly and Lochbaum [14] with computer-music accompaniment by Mathews [15]. To get to a DWG voice model, it was only necessary to "erase" scattering junctions over sufficiently uniform stretches of vocal tract, such as from the larynx to the velum.

It is surprising that ladder filters were not routinely being used as computationally efficient and robust physical models of wave propagation. Instead, their use was normally abstracted away from their physical underpinnings, leaving single-input single-output digital filters, insufficient for use as physical modeling building blocks [16]. The field of signal processing became more rooted in applied math, abandoning its origins in physical analog circuits for the most part, with one exception being *wave digital filters* (§1.3).

### 1.1 Digital Waveguide Synthesis

In the original DWG paper [6], it was suggested that

"waveguide structures can provide accurate models of coupled vibrating strings, wind instruments, reed instruments, and many other physical systems. ... coupled to a nonlinear 'excitation element,' such as a reed, bow, switching air-jet, or lips [17]."

Subsequent work in that direction led to 'Efficient Simulation of the Reed-Bore and Bow-String Mechanisms' [18] the following year.

Coincidentally, Stanford's frequency-modulation (FM) synthesis patent was nearing expiration, so the Stanford Office of Technology and Licensing sent the patent's author David Lovejoy to CCRMA to obtain all the latest papers. A patent application was submitted entitled 'Digital Signal Processing Using Closed Waveguide Networks.' It was initially rejected due to its emphasis on 'closed networks of lossless digital waveguides' (for reverberator prototypes) and the patent office's rule that all proposed 'perpetual-motion machines' lack *utility* because they are physically impossible to realize [19]. After a lot of back and forth, the patent was issued [20], and so lossless reverberator prototypes apparently became the first patented virtual perpetual-motion invention in the United States (at least when roundoff-error feedback is used, which was done in some cases).

Yamaha began developing waveguide synthesis right away, and in 1989 signed a non-exclusive license with Stanford. Yamaha's Director of Engineering was present at the presentation of [18], as was Bob Moog, who wrote a newsletter article about it. In 1994, the Yamaha VL1 *Virtual Lead* synthesizer [21] was intro-duced, based on digital waveguide modeling. The time was ripe for this approach to physical modeling synthesis, as computational power at the time could not afford a full finite-difference-scheme approach as would be typical in computational physics.

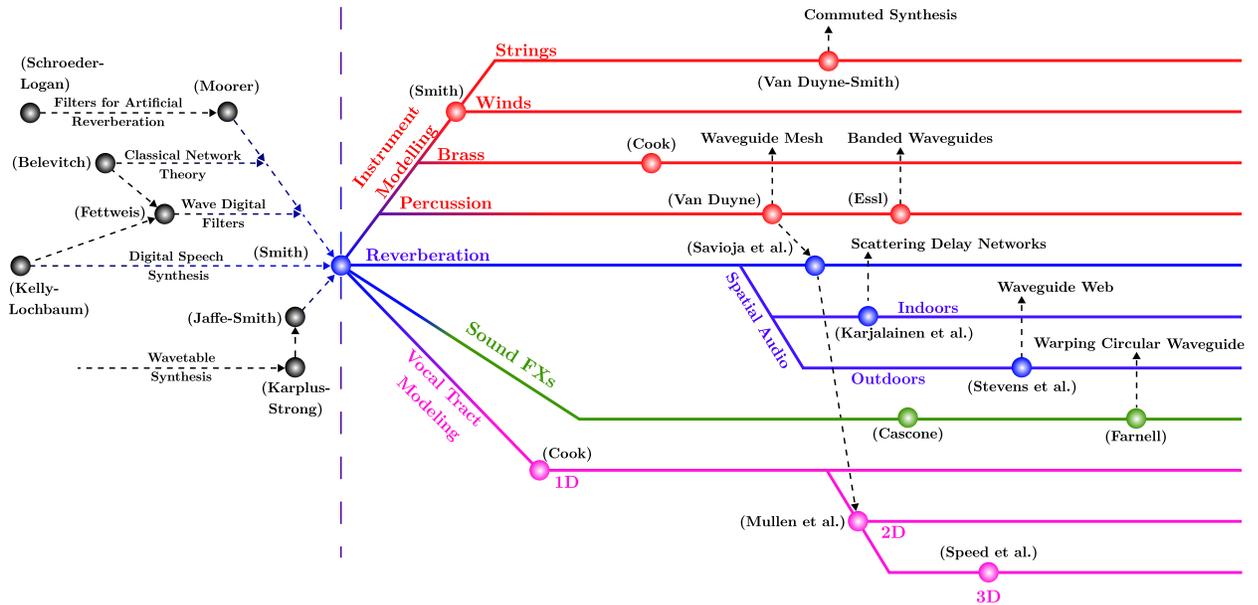

Fig. 1. The lineage of digital waveguide modeling.

**1.2 Relation to the Karplus-Strong Algorithm**

The Karplus-Strong Algorithm (KSA) [22], introduced in 1983, is often described as physical modeling synthesis. But that's not how it was understood at the time. Instead, KSA was seen as *wavetable synthesis* that modified its wavetable on each pass through the table. For plucked string sounds, each sample was replaced by the average of itself and the previous sample, requiring no multiplies. Pitch control was achieved by setting the wavetable length to $N = f_s/f_0 - 1/2$ samples, where $f_s$ and $f_0$ are the sampling and fundamental frequency respectively. As Kevin Karplus and Alex Strong noted [22]:

> 'interesting to compare the theoretical analysis of our synthesis technique [calculating overtone decay times] with an existing analysis of a guitar, lute, mandolin, or other string instrument. Unfortunately, we could not find a published analysis and did not have the tools to perform our own analysis. Some previous work has been done using physical models for synthesizing string sounds (Hiller and Ruiz 1971 [23]), but these models do not help to explain the high quality of the sound produced with our technique. … interesting to convert the recurrence relations to differential equations, and to assign a physical interpretation.''

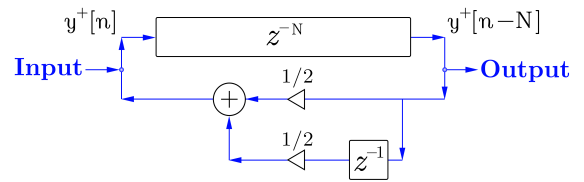

Fig. 2. A digital signal-processing diagram of the Karplus-Strong Digitar algorithm, with additive input added.

Strong played in a string quartet with Music doctoral student David Jaffe, to whom he demonstrated the 'Digitar' algorithm. Jaffe brought it to EE PhD student Smith to see about making it play in tune, adding dynamic level variations, and so on. Smith translated the KSA to a signal processing formulation and recast it as a delay line with explicit additive input, as seen in Fig. 2 above. It was then straightforward to develop Jaffe's desired signal-processing extensions [24].

The tuning problem was solved by calculating the phase delay of the *attenuation filter* [25, p. 170] (now more general than the two-point average seen in Fig. 2), and compensating with delay-line interpolation—initially first-order allpass interpolation [25, p. 175] and later Lagrange polynomial interpolation [26, 27]. Smith recognized the filter's equivalence to point-to-point transfer-functions on an ideal string [25, pp. 158-194]. Impulse responses representing the vibrating-string 'history' had already been introduced for bowed strings [28], but no one had suggested a round-trip string impulse-response as simple and sparse as that of a delay-line feeding back to its input through a two-point average. These insights gave rise to the pick-position illusion added in the Extended Karplus Strong algorithm [24]. Modulating the attenuation filter explicitly alters the decay and timbre. As we will see in §4.2, multiple attenuation filter designs have been proposed.

This was not yet digital waveguide modeling, however, because all signal paths involved *unidirectional* transfer functions with no notion of physical parameters such as string tension, mass, and/or damping. DWGs, in contrast, were conceptually *bidirectional* delay lines associated with a real, positive *wave impedance* calculated as the square root of string tension divided by mass density [6, 16]. To avoid terminology collisions with ladder and lattice digital filters, *delay line* here means *two or more* samples of delay. Typically, the final DWG model is simplified to reduce computational complexity while maintaining the physical interpretation. As an example, KSA was derived as a simplified physical model of a lossy vibrating string, randomly struck and plucked at each sample across its length [16]. That interpretation becomes more transparent when the order of the string and its resonator are swapped, or 'commuted'. This is shown in §2.5 ("Commuted Synthesis"), where KSA emerges as a simplified waveguide model.

### 1.3 Relation to Wave Digital Filters

While DWGs hold *sampled traveling waves*, wave digital filters hold *bilinearly transformed traveling waves*. Both use traveling-wave *scattering theory*, which has its origins in 19th-century optics (Fresnel 1823), acoustics (Lord Rayleigh 1871), and electrical engineering (Heaviside 1886). Lord Rayleigh is said to have first observed the proportionality of pressure and velocity in a traveling wave [29]. The scattering formulation for circuit-theory ports was introduced by Belevitch [30], and "$s$ parameters" have been used since as optional port descriptions by microwave engineers, physicists, and others. This formulation was taken to the digital domain by Fettweis [31, 32] who coined the term *wave digital filter* (WDF).

The *bilinear transform*, introduced by Tustin [33] and Steiglitz [34, 35], maps continuous voltage waves to discrete time in WDFs. Unlike sampling, bilinear transforms introduce *no aliasing*. The price for this is *frequency warping* according to $\omega_c \propto \tan(\omega_d T/2)$, where $\omega_c$ and $\omega_d$ denote continuous- and discrete-time radian frequency, respectively, and $T$ denotes the sampling interval in seconds. The WDF port impedances are chosen such that capacitors and inductors become unit delays [16], thus preserving system order.

Both DWGs and WDFs simulate traveling-wave components in known wave impedances. So they are readily interconnected despite having divergent frequency axes at high frequencies. WDFs are often used to model *lumped* systems involving masses, springs, and friction which are direct analogs of inductors, capacitors, and resistors. Examples include the *wave*

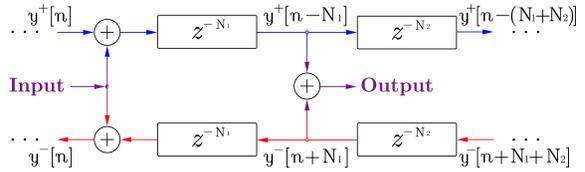

Fig. 3. Discrete-time simulation of ideal, lossless wave propagation using a DWG. Red lines indicate right-going traveling waves, while blue stands for left-going traveling waves.

*digital piano hammer* [36, 37], and the *wave digital tonehole model* [38]. WDFs are also used to create real-time digital models of nonlinear audio effects [39, 40] such as diode clippers [41] and tube amplifiers [42, 43]. DWGs efficiently implement *distributed* wave propagation media, such as vibrating strings, acoustic tubes and tonal percussion. WDFs, in contrast, digitize *lumped* networks, such as mass-spring systems or RLC circuits.

## 2 FOUNDATIONS

A digital waveguide is a *bidirectional delay line* [16] associated with real, positive *wave impedance $R$* which is given by the ratio of force to velocity in a traveling wave [16, 44, 45]. In practice, two separate delay lines may be used, one containing 'left-going' traveling-wave components and the other 'right-going' waves. Summing a signal into a DWG corresponds to superimposing a disturbance at one point. The original state is unaffected; the input signal enters the waveguide in superposition with the current traveling wave. This is depicted in Fig. 3, as a DWG with input at $n = 0$ and output at $n = N_1$.

The *sum* of left- and right-going components yields the physical wave variable such as velocity, pressure, force, current or voltage [16]. In many applications one can use a single delay line in which the first half corresponds to traveling waves to the right, say, while the second half contains left-going samples, thereby implicitly modeling a non-inverting reflection at the midpoint of the single shared delay line.

A DWG models a 1D traveling-wave medium, such as an ideal vibrating string or acoustic tube. For frequencies up to Nyquist limit, the modeling is exact in the ideal case because there are no losses, dispersion, or nonlinearities in an ideal string or acoustic tube.

### 2.1 Vibrating Strings

For vibrating strings, the simulated variables are typically velocity, force, displacement, or the acceleration at each sample point along the string. A typical choice is *transverse string velocity*, as measured directly by a magnetic pickup in an electric guitar. The scattering formulas are unchanged for displacement and acceleration, and force waves are similar with some sign flips.

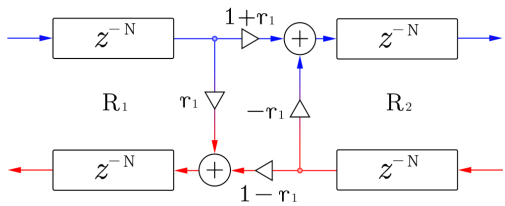

Fig. 4. Scattering junction.

Let $v[n,m]$ denote the physical velocity at temporal sample $n = 0,1,2,\ldots$, and spatial sample $m = 0,1,2,\ldots,M-1$. So our digital string is simulated at $M$ uniformly spaced samples over discrete time. For more physical time and position notation, we can define $v_p[nT,mX] = v[n,m]$ where $T$ is the temporal sampling interval in seconds, and $X$ is the spatial sampling interval in meters. Then in terms of traveling-wave components we have [16]

$$\begin{aligned} v[n,m] &= v_+[n,m] + v_-[n,m] \\ &= v_+[n-m,0] + v_-[n+m,0] \\ &= v^+[n-m] + v^-[n+m], \end{aligned} \quad (1)$$

where $v_+[n,m]$ denotes the right-going and $v_-[n,m]$ the left-going traveling-wave components at time $n$ and position $m$. We define the simpler $v^+[n-m]$ and $v^-[n+m]$, which only depend on time, making use of the fact that a spatial shift corresponds to a time delay or advance in a delay line.

*Wave impedance* of an ideal string is given by the square root of the string *tension $K$* times the *linear mass density* $\mu$, or $R = \sqrt{K\mu}$ [16]. The wave impedance is thus the geometric mean of the two resistances to motion: a spring constant $K$ and inertial mass $\mu$. Similarly, *wave speed* is given by $c = \sqrt{K/\mu}$.

Traveling velocity waves can be easily converted to traveling force waves using wave impedance $R$:

$$f^+[n] = R v^+[n], \quad (2)$$
$$f^-[n] = -R v^-[n]. \quad (3)$$

*Input* to a DWG is usually added equally to left- and right-going directions [16]. This is often accompanied by left and right *reflections* ("scattering") at the input point, because physical inputs will change the wave impedance seen at that point in the waveguide [16].

## 2.2 Wave Impedance and Scattering

Digital Waveguides can be interconnected to form *Digital Waveguide Networks* by joining them at *scattering junctions*. When a waveguide at impedance $R_1$ connects to a waveguide at a different impedance $R_2$, *scattering* occurs, meaning partial *reflection* and *transmission* in both directions. Consider a right-going traveling-wave $v_1^+[n-M_1]$ at wave impedance $R_1$ encountering the impedance $R_2$, which creates a *scattering junction* at the $R_1 : R_2$ interface, as shown in Fig. 4. The incident wave splits into a reflection $v_1^-[n-M_1]$ to the left, and transmission $v_2^+(n)$ to the right. Given no second incoming wave from the right, $v_2^-(n) = 0$. Since all waves on the left at the junction have time index $n - M_1$, and all waves on the right at the junction have time index $n$, we can omit the indices for notational simplicity. So we know $v_1^+$, want to calculate $v_1^-$ and $v_2^+$, and have two constraints:

1. By Newton's third law, the transverse forces must sum to zero at the junction:
$$[f_1^+ + f_1^-] + [f_2^+ + f_2^-] = 0 \quad (4)$$
$$\Rightarrow \quad \frac{v_1^+ - v_1^-}{R_1} + \frac{v_2^+ - v_2^-}{R_2} = 0. \quad (5)$$

2. If the string does not break, then the string endpoint velocities are equal at the junction:
$$v_J = v_1^+ + v_1^- = v_2^+ + v_2^- = v_2^+. \quad (6)$$

These physical constraints are counterparts of Kirchhoff's voltage and current laws in circuit analysis. We can solve for the outgoing waves:

$$v_1^- = v_1^+ \frac{R_1 - R_2}{R_1 + R_2}, \quad v_2^+ = v_1^+ \frac{2R_1}{R_1 + R_2}. \quad (7)$$

In terms of the *force reflection coefficient*

$$r_1 = \frac{R_2 - R_1}{R_1 + R_2}, \quad (8)$$

we can write the complete *scattering equations* for this *series* junction of Fig. 4:

$$f_1^- = r_1 f_1^+, \quad f_2^+ = [1 + r_1] f_1^+, \quad (9)$$
$$v_1^- = -r_1 v_1^+, \quad v_2^+ = [1 - r_1] v_1^+. \quad (10)$$

It is a 'series' junction because there is a common velocity, while forces sum to zero. By contrast, a 'parallel' junction has a common force, and velocities sum to zero.

As $R_2 \to \infty$ (rigid termination), $v_1^- = -v_1^+$ (an inverting reflection) and $v_2^+ = 0$ (no transmission). Similarly, when $R_2 = 0$ (a 'free end' which can be imagined as a frictionless guide-pole) we get $v_1^- = v_1^+$ (non-inverting reflection) and $v_2^+ = 2v_1^+$ (giving velocity continuity across the junction). Finally, when $R_2 = R_1$, we get $v_1^- = 0$ and $v_2^+ = v_1^+$ (no scattering).

Note that power is conserved, i.e., the scattering is *lossless*. The incident traveling power is $f_1^+ v_1^+$, and the outgoing power after scattering is $f_2^+ v_2^+ - f_1^- v_1^-$. Calculation is easy in terms of the reflection coefficient:

$$\begin{aligned} f_2^+ v_2^+ - f_1^- v_1^- &= [1+r_1] f_1^+ [1-r_1] v_1^+ \\ &\quad - r_1 f_1^+ [-r_1 v_1^+] \\ &= [1-r_1^2] f_1^+ v_1^+ + r_1^2 f_1^+ v_1^+ \\ &= f_1^+ v_1^+. \end{aligned} \quad (11)$$

## 2.3 Acoustic Tubes

The vocal tract in speech is often modeled as a piecewise-cylindrical acoustic tube [16], as shown in

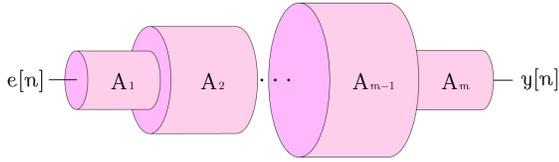

Fig. 5. Kelly-Lochbaum piecewise cylindrical tube model of the vocal tract, where $A$ indicates the cross-sectional area of each corresponding section.

Fig. 5. In this model, the joining planes between different sections are scattering junctions (as in Fig. 4) in the sampled traveling-wave formulation.

Signal samples are typically acoustic *pressure*, and wave impedance in the $m$th tube section (pressure divided by volume-velocity) is given by $R_m = \rho c / A_m$, where $\rho$ is air mass density (kg/m$^3$), $c$ is the speed of sound (m/s), and $A_m$ is the cross-sectional area of the $m$th tube section (m$^2$).

The scattering formulas at the junction of cylindrical segments $m$ and $m+1$ can be written in terms of the *reflection coefficient $r_m$*, defined by

$$r_m = \frac{R_{m+1} - R_m}{R_{m+1} + R_m} = \frac{A_m - A_{m+1}}{A_m + A_{m+1}}. \qquad (12)$$

Thus, the reflection coefficient $r_m$ depends on the *area ratio* $A_m/A_{m+1} = R_{m+1}/R_m$ between sections. Using this notation for the string gives $v_1^- = -r_1 v_1^+$, where the minus sign arises because we used velocity instead of pressure as the wave variable.

## 2.4 Waveguide String Models

A basic waveguide string model is based on a pair of delay lines, where the upper delay line simulates the wave propagation from the bridge to the nut and the lower delay line propagates in the opposite direction [16]. All samples move one position each sampling interval by incrementing (or decrementing) a pointer to a string endpoint in the circular-buffer implementation of the delay line. At each end of the string, the sampled wave reflects backward, optionally transmitting as well. In the lossless case, the reflection coefficient is $-1$ for velocity (or displacement) at a rigid termination. Ideal reflection at the bridge end means that the incoming velocity sample arriving along the lower rail gets its sign inverted as it jumps to the upper rail, where it travels toward the other end. However, oscillations never decay in this undamped model, so a periodic signal is produced after any excitation.

A more realistic waveguide string model has lowpass filters at one or both ends, which model the frequency-dependent propagation losses along the string as well as the reflection from the string end [46–48]. If the total gain of the round-trip filtering is less than 1 at all frequencies, the sample amplitudes decay gradually, which corresponds to the decay of the string vibration.

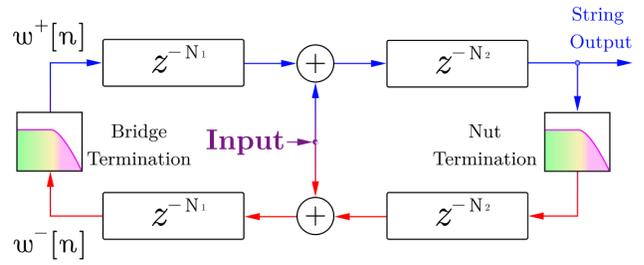

Fig. 6. Discrete simulation of a string using filters as bridge and nut terminations. In the rigid case, the filters become a -1 multiplication for velocity or displacement waves.

## 2.5 Commuted Synthesis

Since basic waveguide string models consist of linear, time-invariant elements, the elements can be reordered in series without changing the transfer function. So the two delay lines of a DWG may be combined into one doubly long delay line. Similarly, the two reflection filters can be commuted to the same point in the delay-loop and combined in a single filter realizing all the filtering necessary in the whole delay loop. We refer to this generic structure, similar to that of the KSA (see §1.2), as a *filtered delay loop* (FDL) [16, 49].

In a plucked-string instrument, the output signal $y[n]$ can be expressed as a cascade of convolutions:

$$y[n] = e[n] * h[n] * b[n], \qquad (13)$$

where $e[n]$ is the *excitation signal* (e.g., the pluck) and $h[n]$ and $b[n]$ stand for the impulse response of the string and instrument body respectively. Since convolution is commutative, these elements can be reordered without changing the output.

This affords great computational savings for strings attached to resonators [50]. Instead of plucking a string and driving a high-order body resonator, such as a guitar body or piano soundboard [51, 52], one can pluck the resonator and feed that into the string [46, 50, 53]. The savings comes from storing only the impulse response (or 'pluck response') of the resonator $e[n] * b[n]$. So the large digital resonator models needed for acoustic stringed instruments (hundreds or thousands of poles and zeros occurring in the audio range) are replaced by a basic wavetable lookup and playback through a small filter that can model variations in plucking details. This also applies to bowed strings modeled as periodically plucked strings [50, 54, 55].

## 2.6 Digital Waveguide Mesh

The *digital waveguide mesh* (DWM) simulates a membrane or volume using a lattice of 1D strings, analogous in 2D to a tennis racket [56]. A 2D membrane is thus approximated by a mesh of intersecting 1D waveguides, as depicted in Fig. 7. Each intersection forms a four-port scattering junction. Typically, all wave impedances are equal so that the junction velocity $v_J$ can be computed using *no multiplies* as $v_J =$ $\frac{1}{2}\sum_{i=1}^{4} v_i^+$; i.e., sum of four incoming waves $v_i^+$ divided by 2. Each outgoing wave is then $v_i^- = v_J - v_i^+$ for a total of seven additions and one bit-shift (in fixed-point). Lossless wave propagation is exact along the diagonals but *dispersive* along horizontal and vertical directions [57]. To reduce dispersion at the cost of multiplies, the *triangular mesh* is a good choice [58–60].

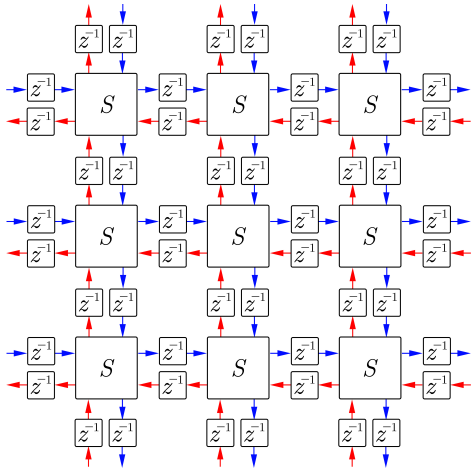

Fig. 7. 2D rectilinear waveguide mesh. $S$ stands for scattering junctions.

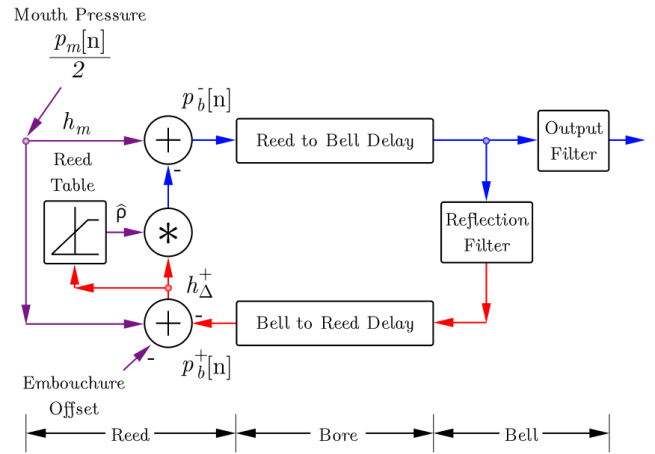

Fig. 8. Digital waveguide model of a single-reed woodwind instrument.

For multiply-free scattering in 3D, the *tetrahedral mesh* geometry is preferred [61, 62]. Later mesh developments include acoustic room simulation [63–67], the bilinearly deinterpolated waveguide mesh [68, 69], high-frequency mode modeling [70, 71], the *frequency-warped* waveguide mesh [59, 72, 73], violin body modeling [74], boundary conditions [75], quadratic residue diffusers [76], extensions to higher dimensions [77], rigid and diffuse boundary modeling [78, 79], and vocal-tract modeling [80–84].

## 3 ADVANCEMENTS

### 3.1 Instrument Modeling

#### 3.1.1 Winds

The first waveguide synthesis instrument was the clarinet [18] (Fig. 8). Pressure waves $p$ propagate through the bore and interact with the mouthpiece, where subscripts $m$ and $b$ denote mouth and bore, and superscripts $+$ and $-$ indicate right- and left-going waves. Half-pressures $h = p/2$ are used, where $h_\Delta^+$ across the reed corresponds to the pressure drop driving nonlinear flow. The clarinet reed was modeled as a massless spring flap [85, 86], acting as a nonlinear reflection coefficient $\hat{\rho}$ at the tube end—mapping the pressure difference across the reed to the resulting airflow—and a nonlinear variable gate for airflow into the tube.

The initial implementation of the reed-bore reflection-coefficient (a function of the incoming pressure waves from both sides, and the player's embouchure) was a linearly interpolated table-lookup. This pre-solved nonlinear scattering junction can be regarded as a 1D version of the *K Method* [87], also applicable to root-node scattering in nonlinear Wave Digital Filters (WDF) [39]. Later implementations included piecewise polynomial spline approximations [88, 89]. Some reed models also included mass [85, 90] which cannot be pre-solved, so a few Newton iterations or the like are used to solve for the nonlinear junction outputs.

The second waveguide wind-instrument model was the *flute* [91–94]. The mouthpiece excitation was modeled as an air jet [17] that swung in and out of the tube based on air-jet dynamics and the traveling pressure wave in the bore at the mouthpiece. Since the flute mouthpiece is in the interior of the bore, two waveguides are used, one on each side of the excitation.

The first real-time waveguide saxophone was in the Yamaha VL-1 synthesizer [21]. It is said to have used Benade's two-cylinder model for the conical bore [95–97]. Conical bores become interesting near the tip, which behaves like an open end for pressure [98]. Another oddity is that the impulse-response of an increase in the conical taper-angle has a growing exponential component [99]. Woodwind toneholes were modeled as variable scattering junctions in the digital waveguide bore [100, 101], and a wave digital tonehole was developed [38].

#### 3.1.2 Strings

Waveguide synthesis for strings began with simplified bowed strings [18], which affixed a bow-string interaction model [17, 28] to a one-branch digital waveguide network [6], as shown in Fig. 9. In the model, velocity-wave components $v$ travel along the string, with subscripts $s,r$ and $s,l$ referring to string segments to the right and left of the bow respectively. This divided the string into two waveguides joined by a nonlinear two-port scattering junction, where the bow-string interaction is governed by the differential velocity $v_\Delta^+$ (bow velocity minus incoming string velocity) and the nonlinear reflection function $\hat{\rho}$, which depends on bow force and velocity.

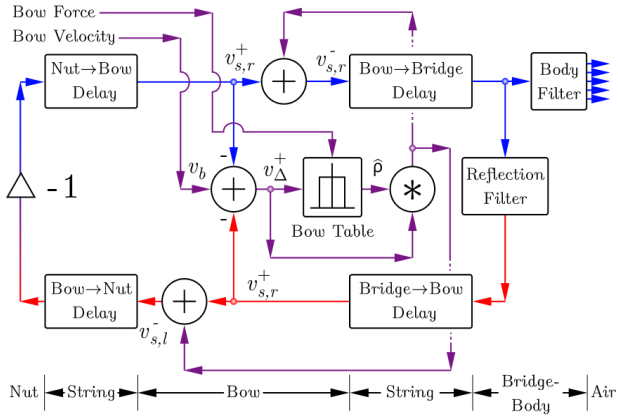

Fig. 9. Digital waveguide model of a bowed string.

Bow position is varied by adjusting the relative lengths of the left and right delay lines, with interpolation providing continuous control. Assuming the bow-hairs present only a nonlinear frictional force (i.e., neglecting bow-hair traveling waves), a table-lookup or piecewise polynomial solution is obtained for the instantaneous reflection and transmission coefficients [18]. Later models introduced friction-loop hysteresis and thermodynamic models of rosin friction [102–106].

DWGs have been used for modeling various plucked string instruments, such as the acoustic guitar [48, 53, 107], the Finnish kantele [108, 109], and the geomungo, a traditional Korean instrument [110]. To capture the unusually wide vibrato effect of the geomungo, digital waveguide synthesis with a time-varying attenuation filter was used [110]. Some keyboard instruments, which are also based on vibrating steel strings, have been emulated with DWGs, such the piano [111–114], the harpsichord [115], and the Clavinet [116]. The electric guitar is a popular target for DWG synthesis [117–120].

### 3.1.3 Brasses

Waveguide *brass* instruments began with the trombone [88]. In brass models, the lips are typically modeled as one or two mass-spring elements driving the waveguide tube. Flared bells inspired applications of Sturm-Liouville theory [121], and the most efficient known filter structure for the flared bell is truncated infinite-impulse-response filters [99].

### 3.1.4 Percussion

*Tonal* percussion, such as glockenspiel and triangle, are straightforward to model as DWGs, being composed of $\approx$ 1D metal rods. *Membrane* percussion, such as tympani and other drums, can use a 2D waveguide mesh for their membranes, and a 3D mesh for the enclosed air cavity. Another approach to membranes, which has been applied to *cymbals*, is the *banded waveguides* technique, which identifies *closed wavetrains* modeled by 1D digital waveguide loops [122–124].

### 3.1.5 The Synthesis Tool Kit (STK) and Faust

The Synthesis Tool Kit (STK) [89] is a major software repository of example DWG instruments. An extended subset of the STK examples was ported to the Faust distribution as well [125]. The Web Audio examples for this review include ports of STK examples.

In addition to ports of the original STK classes, the Faust Libraries [126] provide many additional synthesis and effects algorithms using digital waveguides. The term 'waveguide' appears 55 times, as of this writing, spanning the libraries for oscillators, noises, filters, physical models, and miscellaneous effects.

### 3.2 Vocal Tract Modeling

The Kelly-Lochbaum model [14] is thought to be the first digital scattering-based model of the vocal tract, and influenced ladder and lattice filter development for linear prediction of speech [13]. Smith has provided a detailed derivation of wave impedance and scattering theory starting with D'Alembert's wave equation [16].

DWG modeling of *singing human voice*, introduced by Cook [127], can be considered a generalization and extension of the Kelly-Lochbaum model. A real-time interactive singing-voice model called SPASM ("Singing Physical Articulatory Synthesis Model") [2, 128] included a nasal tract branching off from the vocal tract through a variable three way velum/pharynx junction, turbulence noise injectable at any point along the vocal tract, variable glottal and lip reflection coefficients, and separate outputs for neck, lip, and nose radiation. A related interactive implementation called 'Pink Trombone' is available online [129].

Later advancements used DWMs (§2.6) for the vocal tract. Early 1D vs 2D comparisons and bandwidth control [130, 131] revealed that 2D relaxation of the plane-wave assumption introduces lateral modes and provides approximately linear formant-bandwidth control across vowels [132]. Real-time dynamic articulations in a fixed rectangular 2D mesh using impedance mapping were then introduced [133], and 3D meshes that match measured transfer functions for tract analogs demonstrated close acoustic match to physical analogs over a wide frequency band [81, 134]. Recent results use MRI-derived geometry and time-varying impedance maps to synthesize diphthongs with improved formant accuracy and naturalness [135], and voiced stop consonants via controlled occlusions in a 3D DWM [84].

Beyond the human laryngeal tract, waveguide principles have been applied to other vocal organs. In songbirds, the syrinx—a bifurcated airway with vibrating membranes—has been modeled using classical waveguide synthesis [136–138]. Mammalian vocal tracts were implemented using a waveguide filter model in [139]. Subjective tests showed that it was effective in reproducing harsh, spectrally dense sounds such as a lion's roar or a wolf's growl.

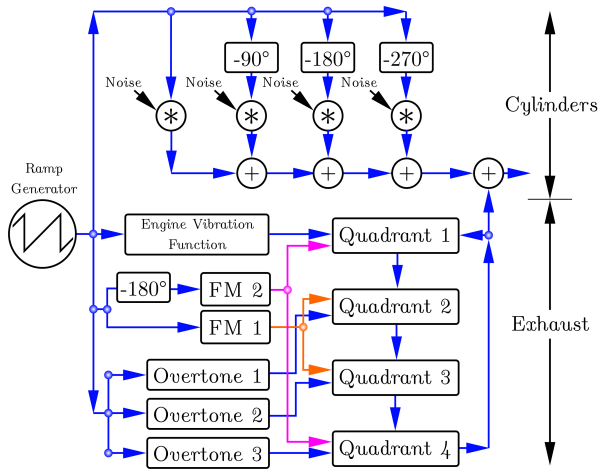

Fig. 10. Four-cylinder engine model using digital waveguides. Pink and orange arrows indicate FM control signals (not audio inputs).

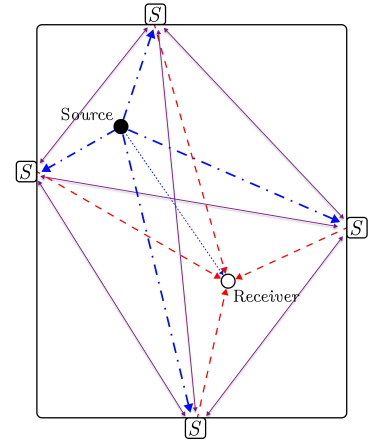

Fig. 11. Scattering Delay Network (SDN). Wall nodes ($S$), bidirectional delay lines (purple), source paths (dash-dotted blue), receiver paths (dashed red), direct path (dotted blue).

### 3.3 Sound Effects

DWGs are also used to simulate various mechanical devices, thus creating realistic sound effects for use in creative media and computational modeling. The most common example of this is in simulating vehicle engine sounds, where DWGs are often used for engine cylinders and exhaust pipes. The first example was perhaps given by Cascone et al. [140], for which a Faust implementation is available [141].

Farnell introduced DWGs with anti-phase frequency-modulated delays for simulating engine exhausts, controlled by a ramp generator mimicking mechanical inertia [4]. These delay lines are split into four *quadrants*, with some expanding and others compressing while maintaining constant length, as shown in Fig. 10. He coined this the *Warping Circular Waveguide* [4]. The Engine Vibration Function applies a parabolic waveshaper to model low-frequency body resonance, while *Overtone generators* inject bursts of higher-frequency content at specific positions, simulating mechanical transients (valve clicks, combustion noise). Separately, cylinders are simulated by producing pulses from the phase-shifted ramp, multiplied by noise for jitter. This approach was further developed by Baldan et al. who described car engine sound synthesis with multiple DWGs [142]. Two subsequent works incorporated Baldan's implementation into soundscape synthesis of traffic sounds [143, 144].

Farnell's use of DWGs, though informal and often heuristic, was wide in scope. DWGs were incorporated into models of not just automotive engines, but models of old telephones, impacts on windows, switches and levers, and helicopters [4].

Other applications of DWGs for generating sound effects are of note. Oksanen et al. modeled the sound of a jackhammer using parallel waveguides for longitudinal and transversal vibrations [145]. Use of banded DWGs for generating the *twang* sound of a ruler being pushed while it is hung over a desk was described by Selfridge et al. [146], though they opted for a modal synthesis approach in their implementation.

In [147] and references therein, Keenan and Pauletto simulated the sound of an acoustic wind machine, which is a mechanical device that in turn simulates wind sounds. The machine has a slatted cylinder whose slats rub against a cloth as the cylinder moves. The resultant friction produces a wind-like sound. A digital waveguide in series with low-pass and all-pass filters was used to simulate dispersion of the friction sound through one side of the cloth. Use of DWGs for real-time, controllable sound effect generation is discussed and put in the wider context of *Procedural Audio* in a review paper by Menexopoulos et al. [148].

### 3.4 Reverberation

The first *digital audio effect* using a 1D waveguide network was an "invertible waveguide reverberator", showcased in the finale of "Invisible Cities" ("City of Reflection" movement) by Michael M. McNabb [149]. In this striking effect, a saxophone passage is scattered to "sonic dust" by a closed, lossless network, and then that network runs in reverse to congeal the dust back to the original sax passage. Despite this early artistic success, there was little widespread application to reverberation until the introduction of the *waveguide mesh* (see §2.6). Widely adopted Feedback delay networks [150] can be described as DWGs having one scattering junction, but they were not derived that way.

Physical modeling networks of 1D waveguides for spatial audio awaited the advent of *scattering delay networks* [151, 152], illustrated in Fig. 11. In the SDN, the source drives the wall nodes via unmodified, unidirectional paths, while the receiver collects scattered components via unidirectional paths. The *direct* sound from source to receiver follows the free-field direct path. The *first-order reflection* paths from source to receiver pass through the scattering junctions. Four *scattering junctions* denoted $S$ are located at the physical first-order-reflection points along the walls. Six bidirectional delay lines interconnect the four scattering junctions, forming a digital waveguide *network*. The purpose of the network is to simulate the *higher-order reflections* efficiently, albeit with approximation error: While the scattering junctions are located correctly for first-order reflections, they are not generally correct for higher-order reflections.

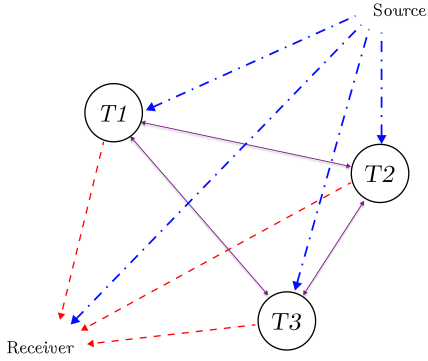

Fig. 12. Treeverb (digital waveguide web). Tree-nodes ($T_1$–$T_3$), bidirectional lines (purple), source paths (blue), receiver paths (red).

Thus, the efficiency of the SDN reverberator is obtained by moving the point of reflection for each higher-order reflection from its exact location along the wall to the location of the first-order reflection on that wall. The main error is that the exact timings of the higher-order reflections are perturbed, but the echo distribution and density are preserved. As established by De Sena et al., *"according to various objective measures of perceptual features, SDN achieves a reverberation quality similar to that of the [theoretically precise Image Method] while having a computational load one to two orders of magnitude lower"* [152].

While SDNs target enclosed room acoustics, similar principles have been adapted for outdoor scenarios. *The Waveguide Web* is a digital reverberation algorithm designed to simulate the acoustics of sparsely reflecting outdoor environments, such as a scene with a few buildings, by using interconnected digital waveguides and scattering junctions to model reflection points [153]. The Waveguide Web extends the Treeverb architecture proposed by Spratt and Abel, which models the acoustics formed by trees in a forest [154]. The Treeverb, shown in Fig. 12, may be interpreted as a 2.5D model of a forest, where the source excites the tree nodes and the receiver gathers the post-scattering components, including the horizontal acoustic path from the source to the receiver point and all scattering paths between the tree nodes. As more trees are added into the system, it becomes computationally inefficient, however.

In the Waveguide Web model proposed by Stevens et al. [153], each node combines the acoustic properties of an entity, such as a building, by storing its reflection properties for the first-order reflection of the source signal and the second-order reflections from the other nodes. For open scenes, such as a courtyard surrounded by a couple of facades, the necessary filters can be measured or simulated. This way, a fairly complicated scene can be simulated with a small number of nodes. The Waveguide Web model accurately characterizes both the first- and the second-order reflections from the nodes [153]. Now the scattering junctions cannot be energy-preserving, since some of the acoustic energy escapes the scene between the buildings. A sky node may be defined to simulate such losses [153]. Nodes for ground reflections could also be incorporated, if they appear to be acoustically important.

## 4 PARAMETRIC OPTIMIZATION IN DIGITAL WAVEGUIDES

Often we wish to calibrate a DWG model to resemble a target sound for realism, or to obtain physical parameters of the sound-producing object, such as bow force or reed pulse. In this sense, parametric optimization in DWGs can be viewed as an architecture-specific instance of the broader synthesizer sound-matching problem overviewed by Shier [189].

Over the last four decades, multiple strategies have been applied to DWGs, which we organize chronologically. *Physics-based* approaches rely on analytical relations and measurements. *Filter-design* methods approximate the DWG by placing poles, zeros or coefficients of its transfer function $H(z)$. *System identification* methods pre-compute input-output mappings for online inference (for an early review of filter design and system identification, see [25]). *Genetic* approaches perform random searches where successful parameter sets survive and others are discarded. Finally, *neural* methods optimize parameters by backpropagating loss-gradients through differentiable neural operations. Neural approaches can be further divided into *black-box* methods, where the synthesizer's inner workings are unknown, and *white-box* methods, where they are known and explicitly modeled.

### 4.1 Physics-based

Resonating acoustic bodies are often modeled via *modal expansion* to obtain an impulse response in the form of a sum of exponentially damped sinusoids,

$$x(t) \approx \sum_{n=1}^{N} A_n e^{-\alpha_n t} \cos(\omega_n t + \phi_n), \qquad (14)$$

and we estimate the modal parameter set $(A_n, \omega_n, \alpha_n)$ with some spectral method to design the attenuation filter of a FDL system. Guillemain et al. [155] and

Table 1. Optimization Methods for Digital Waveguides.

| Paradigm | DWG | References | |
|---|---|---|---|
| Physics-Based | Filtered Delay Loop | [155, 156] | Analytical Relations |
| | Clarinet | [157, 158] | Real-Life Measurements |
| Filter Design | Filtered Delay Loop | [159] | Prony |
| | | [46, 47, 160, 161] | Iterative Weighted Error |
| | | [113, 162, 163] | Steiglitz–McBride |
| System Identification | Kelly–Lochbaum | [164] | Nearest Neighbor |
| | Reed–Bore | [165] | Cosine-Similarity |
| | Bowed String | [166] | Gaussian Mixture Model |
| | Avian Vocal Tract | [138] | Maximum Likelihood |
| Genetic | Flute | [93] | Spectral Fitness |
| | Filtered Delay Loop | [167] | |
| | 2D Vocal Tract DWM | [168] | Perceptually-Informed Fitness |
| | Pink Trombone | [169, 170] | Covariance Matrix Adaptation Evolution Strategy |
| Neural — Black-Box | String | [171] | Distal Teacher |
| | Filtered Delay Loop | [172] | Neural Network with Perceptual Mapping |
| | Reed–Bore | [173, 174] | Neural Exciter Models |
| | Flue-Organ | [175–177] | Convolutional Neural Network |
| | Bowed-String | [178] | Adversarial Autoencoder |
| | Pink Trombone | [179] | Transformer |
| Neural — White-Box | Guqin, Plucked-Strings... | [180–184] | Scattering Recurrent Networks |
| | Filtered Delay Loop | [185, 186] | |
| | Pink Trombone | [187] | Differentiable Digital Signal Processing |
| | Scattering Delay Network | [188] | |

Ystad [156] match $(\omega_n, \alpha_n)$ to viscothermal losses in tubes and stiffness-induced dispersion on strings. However, they report this method does not guarantee realism [155, 156]. In practice, physics-based approaches are often hybridized with other methods due to their potential to simplify optimization by fixing variables: Smyth and Abel, for example, anchor a clarinet waveguide with measured bell reflection filter to obtain its exciting *reed pulse* [157, 158].

## 4.2 Filter Design

Alternatively, $(A_n, \omega_n, \alpha_n)$ can be used to derive the location of poles and zeros of the transfer function $H(z)$ through some filter-design method [47, 160]. The strategies used are summarized in Table 1. Fundamental frequency $f_0$ can be recovered as the lowest detected harmonic [159] or via harmonic spacing [190]; while many works report accurate estimation through autocorrelation [46, 47, 160, 191].

Instrument bodies can be simulated once the string $h[n]$ has been optimized. Välimäki et al. rearrange the terms of Eq. 13 to $y[n] * h^{-1}[n] = e[n] * b[n]$, obtaining the pluck response. This is then stored and reproduced online to simulate guitar, mandolin, kantele, and banjo [46]. Later, Tolonen and Välimäki note the non-harmonic modes of $b[n]$ (for ex. 100-Hz *Helmholtz resonance* in guitars), can bias the filter-design method of $h[n]$ and propose modeling $b[n]$ separately with *Warped*-IIRs, which can be shared along multiple strings [46, 47, 160].

## 4.3 System Identification

Kestian and Smyth [164] use a Kelly–Lochbaum model (see Fig. 5) to precompute a lookup table, mapping area sections to formants. At runtime, nearest-neighbor matches the area from recordings [164]. Similarly, Smyth and Wang [165] estimate saxophone fingering with a 1D reed-bore DWG. The bore values are pre-computed per fingering, choosing the one that maximizes time-domain cosine similarity from bell-mic recordings [165]. Serafin et al. use Linear Predictive Coding and Principal Component Analysis to extract features from a pre-computed bowed-string (see §3.1.2) dataset, then use a Gaussian Mixture Model to learn a probabilistic mapping that predicts bow force and velocity [166]. Similarly, Smyth, Abel, and Smith invert a hybrid DWG-FDTD avian vocal tract model (see §3.2) by pre-computing the power spectra for different bronchial pressure and air volume flow values, used then with maximum-likelihood to estimate their trajectories from new recordings [138].

## 4.4 Genetic

Vuori and Välimäki first used genetic algorithms to optimize a flute model (see §3.1.1), defining a spectral fitness over fundamental and early harmonics [93]. Riionheimo and Välimäki extended this approach by introducing psychoacoustic weighting into the spectral fitness to optimize a plucked-string FDL [167]. Cooper et al. scaled this approach to a 2D-DWM vocal tract (see §3.2) tuning hundreds of spatial parameters [168]. Modern research by Cámara et al. on Pink Trombone optimization demonstrates genetic algorithms remain useful and can compete with Black-Box Neural Network approaches [169]. Furthermore, this strategy can be successfully hybridized; their recent work utilizes Differentiable Digital Signal Processing (DDSP) - see §4.5.2 - to initialize parameters before refining the results with a genetic search [170].

## 4.5 Neural

### 4.5.1 Black-Box

Casey introduced the first neural approach to DWG optimization by inverting a simple bowed-string model through *supervised* training [171]. He fed DWG-synthesized audio $y$ to a neural network and minimized the *parameter loss* $|\hat{\theta} - \theta|$, where $\hat{\theta}$ and $\theta$ are the predicted and known DWG parameters respectively, obtaining a *controller network* able to predict $\hat{\theta}$ from $y$. Additionally, he trained a second network as a *teacher* to predict output waveform $\hat{y}$ by minimizing $|\hat{y} - y|$—which we refer to here as a *synthesizer network*—to help the controller network navigate non-convex optimization landscapes.

Cemgil and Erkut furthered this idea by training a controller network for a plucked-string FDL, with an additional perceptual loss: a subjective mapping of $\theta$ obtained through listening tests [172]. Other work focused on using synthesizer networks to predict $e[n]$ for reed-bore DWGs, expanding the capabilities of DWGs through a data-driven nonlinearity [173, 174].

The introduction of modern neural network architectures has seen substantial improvements in the last decade. Sinclair trained an *adversarial autoencoder* on bowed-string $(\theta, y)$ pairs, obtaining a controllable synthesizer network that mirrors the behavior of the original DWG [178]. Gabrielli et al. demonstrated that *Convolutional Neural Networks* achieve high-quality results when scaled to large datasets, culminating in a patented controller network for a flue-organ DWG [175–177]. Most recently, Xu and Reiss applied *Transformers* to optimize area sections in the Pink Trombone vocal tract model, surpassing neural audio synthesis in reconstruction quality for vowel sounds [179].

### 4.5.2 White-Box

In white-box approaches, the DWG is embedded within the network, allowing backpropagation from output to parameters. Early research by Su, Liang and collaborators exploited the equivalence between Recurrent Neural Networks (RNN) and IIR filters (see [192] for a mathematical derivation). Hence, they designed RNNs whose structure mirrors DWG topologies, where neuron weights act as scattering coefficients [180–184]. However, RNN units need to be cascaded or parallelized to extend beyond first-order, hence complicating the topology of DWGs and impacting computational load at synthesis.

*Differentiable Digital Signal Processing* addresses this by implementing DWGs as native differentiable operations [193]. Parameters can then be optimized through a controller network or directly via gradient descent in a *self-supervised manner*. In the accompanying materials of Hayes et al. [194] an FDL filter's decay is optimized by minimizing the spectral distance between target and synthesized audio at multiple resolutions [185], known as Multi-Scale Spectral Loss ($\mathcal{L}_{MSS}$) [195]. Tablas de Paula et al. extend this approach to optimize pluck position, playing dynamics, timbre and decay separately in a time-domain time-variant implementation [186].

The same technique, gradient descent reducing $\mathcal{L}_{MSS}$, is applied to the Kelly-Lochbaum ladder-filter, where the vocal tract area sections are optimized, scoring higher results in perceptual listening tests than optimization with genetic algorithms, *Particle-Swarm Optimization* and Gabrielli's Black-Box method from §4.5 [187]. Recently, DDSP techniques have been applied to SDNs (see §3.4) to match spectro-temporal features of a target room impulse response. The optimized SDN captures energy decay behavior more realistically than those modeled solely from geometry and tabulated wall absorption coefficients [188].

Table 2. Comparison of Optimization Methods. Nuance is indicated with "~", discussed below.

| Technique | DWG-agnostic | Real Time | Data-Hungry | Quality |
|---|---|---|---|---|
| Physics-Based | Yes~ | No | No | Low |
| Filter-Design | No~ | Yes | No | Medium |
| System ID | Yes | Yes | Yes | Medium |
| Genetic | Yes | No~ | No | High |
| Black-Box | Yes | Yes | Yes | High |
| White-Box | Yes~ | Yes | No~ | High |

## 4.6 Comparisons and Current Challenges

In this review, *physics-based* and *filter-design* approaches have mostly targeted LTI FDLs. However, the former can target arbitrary DWG complexity given derivable or measurable parameters, while the latter supports precise frequency-dependent attenuation [196, 197]. *System Identification* strategies remain useful for deriving input-output mappings, but their generalizability is limited. *Genetic Algorithms* achieve high quality without data requirements, but reconstruction quality depends on population size, affecting real-time performance (Table 2). Modern approaches can significantly accelerate this process [170].

Once trained, neural approaches infer DWG parameters in real-time and generalize to unseen data. Black-box quality depends on data quality and size (*data-hungry*), while white-box methods mitigate this through synthesizer knowledge at the cost of requiring precise differentiable implementations. Both remain limited by loss functions and network capacity.

While parameter losses provide direct gradients, DDSP overrelies on $\mathcal{L}_{MSS}$. For sinusoidal frequency prediction, such as $f_0$, $\mathcal{L}_{MSS}$ yields uninformative gradients [198–201]). To address this limitation, recent work explores alternative loss functions—including combinations of spectral, parameter, and discriminative objectives [202–204], along with hybridization with traditional heuristic searches to refine parameter initialization [170].

Subsequently, when *permutation symmetry* is present, point-based regressors will degrade performance, even when combined with permutation-invariant losses or symmetry-breaking heuristics [205].

Hayes et al. demonstrated that *generative*, *flexible symmetry-equivariant* architectures yields consistent gains in black-box controllers [205].

While generative prediction has entered DDSP, flexible symmetry-equivariant architectures remain absent [206]. In a related development, DDSP has opened a new direction: the combination of physically interpretable-control of physical models with neural audio [207].

# 5 CONCLUSION

Over the past four decades, DWG applications have proliferated, as substantially covered in this review. It is a perfect example of how physically accurate simulations can be perturbed to reduce computations by orders of magnitude while preserving psychoacoustic equivalence. Just as the FM oscillator-pair efficiently approximated a great number of oscillators in additive synthesis, the bidirectional delay lines and scattering junctions of DWG networks greatly reduced the burdensome finite-difference schemes of computational physics for wave propagation. The straightforward calculus of lossless wave scattering—often multiply-free—at sparse points within a general framework of bidirectional delay lines has been found to be intuitively appealing, physically meaningful, and computationally hyperefficient—frequently sounding equivalent to the ear while saving orders of magnitude in computation. Readers are encouraged to explore interactive demonstrations of these principles at the companion website: `https://joshreiss.github.io/digital-waveguides-review/`.

# 6 ACKNOWLEDGMENTS

The authors thank David Marttila for directing us to modern literature on voice modeling. This research was supported by UK Research and Innovation [grant number EP/S022694/1].

## THE AUTHORS

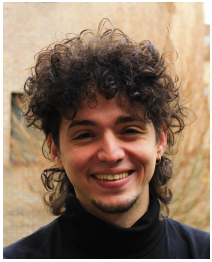 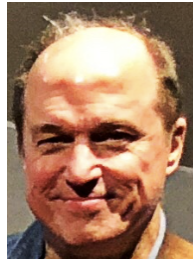 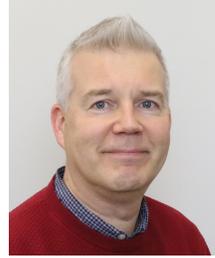 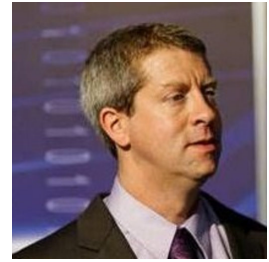

Pablo Tablas de Paula    Julius O. Smith III    Vesa Välimäki    Joshua D. Reiss

Pablo Tablas de Paula is a PhD student with the Centre for Digital Music at Queen Mary University of London. He received his BSc (Hons) degree in Audio Technology from Glasgow Caledonian University in 2024, where he was awarded the JAMES Outstanding Student Award and the Mediaspec Best Honours Project Award. He completed an internship with the Eurorack synthesizer designer Instruō and was part of the R&D team at Blackstar Amplification. Research interests include DDSP, synthesizer sound matching, and digital waveguide synthesis.

•

Julius O. Smith III is Professor Emeritus of Music and by courtesy Electrical Engineering at Stanford University, based at the Center for Computer Research in Music and Acoustics (CCRMA). He received his B.S.E.E. from Rice University in 1975 (Control, Circuits, and Communication), and M.S. and Ph.D. in E.E. from Stanford in 1978 and 1983, respectively. He joined the Stanford Music faculty part-time in 1989. At CCRMA he pursued research and teaching in music and audio signal processing. He was at NeXT Computer, Inc., from 1986 to 1991, tasked with building the Sound and Music Group. He was a founding consultant for Staccato Systems, Shazam, and moForte. For more information, see `http://ccrma.stanford.edu/~jos/`.

•

Vesa Välimäki is a Full Professor of audio signal processing and Vice Dean for research at Aalto University, Espoo, Finland. He received his D.Sc. degree from the Helsinki University of Technology in 1995. His research interests are in signal processing and machine learning for acoustics, audio, and music technology. In 2008, he was Chair of the 11th International Conference on Digital Audio Effects DAFx-08. In 2017, he chaired the 14th International Sound and Music Computing Conference SMC-17. In 2020-2025, Prof. Välimäki was the Editor-in-Chief of the Journal of the Audio Engineering Society and is now the Deputy Editor-in-Chief.

•

Josh Reiss is a Professor with Queen Mary University of London's Centre for Digital Music. He has published more than 200 scientific papers, and co-authored three books. He is a past President and Fellow of the Audio Engineering Society (AES). He has served as an expert witness in major cases related to audio technology and patent litigation. He is Entrepreneur-in-Residence at QMUL, having co-founded four spin-out companies; LandR, Waveshaper AI, RoEx and Nemisindo. His research is primarily on state-of-the-art signal processing and machine learning techniques for sound design and audio production.